\documentclass[pr,aps,superscriptaddress,showpacs]{revtex4}
\usepackage{graphicx}

\newcommand{\dd}{\ensuremath{\text{d}}}

\begin{document}
\title{An extra-heating mechanism in Doppler-cooling experiments}
\author{T. Chaneli\`ere}
\author{J.-L. Meunier}
\author{R. Kaiser}
\author{C. Miniatura}
\author{D. Wilkowski}
\affiliation{\ Institut non lin\'eaire de Nice, UMR 6618 du CNRS,
1361 route de Lucioles, F-06560 Valbonne, France.}
\date{\today{}}

\begin{abstract}

In this paper we experimentally and theoretically investigate
laser cooling of Strontium 88 atoms in one dimensional optical
molasses. In our case, since the optical cooling dipole transition
involves a $J_g=0$ groundstate, no Sisyphus-type mechanisms can
occur. We are thus able to test quantitatively the predictions of
the Doppler-cooling theory. We have found, in agreement with other
similar experiments, that the measured temperatures are
systematically larger than the theoretical predictions. We
quantitatively interpret this discrepancy by taking into
consideration the extra-heating mechanism induced by transverse
spatial intensity fluctuations of the optical molasses.
Experimental data are in good agreement with Monte-Carlo
simulations of our theoretical model. We thus confirm the
important role played by intensity fluctuations in the dynamics of
cooling and for the steady-state regime.

\end{abstract}

\pacs{PACS: 32.80.Pj} \maketitle

\section{Introduction}
Initiated in the mid-seventies, laser cooling, trapping and
manipulation of atoms rapidly became a very successful field of
research \cite{houches90}, culminating twenty years later with the
observation of Bose-Einstein condensation of alkali atoms
\cite{bose}. This success story begun with the seminal Doppler
cooling theory which was designed for two-level systems
\cite{Hansch75,wineland79}. However, the first reliable tests of
the Doppler theory could only be performed at the end of the
eighties. As it turned out, the measured temperatures in the
experiments were \emph{well below} the predicted Doppler values
\cite{lett}. This surprising results indicated that another much
more efficient cooling mechanism was at work in the experiments.
This mechanism, now known as Sisyphus cooling, was identified soon
after \cite{cch89shu89}. The key point was to understand that
internal groundstate degeneracies were opening the way to a new
cooling mechanism based on optical pumping in the presence of
polarization gradients. This cooling mechanism is now the basic
ingredient for most experiments in the field and thus gave the
Doppler theory a more academic status.

Recent laser technological advances however opened the way to
cooling and trapping experiments with earth-alcaline and
rare-earth atoms (Calcium Ca, magnesium Mg, strontium Sr and
Ytterbium Yb). As these atoms exhibit a zero spin groundstate,
Sisyphus cooling is absent. Interest in testing the Doppler theory
to gain better control on achievable temperatures has thus been
renewed. To our knowledge, all experiments on laser cooling of
these atoms \cite{Oates99, loo04, loftus03, yoon04} always
reported much larger temperatures than the Doppler theory
prediction. This is a strong clue in favor of an extra-heating
mechanism still to be clearly identified. Approaching the Doppler
limit in magneto-optical traps (MOT) operating with earth-alcaline
atoms seems a challenge and a better understanding of this heating
mechanism is required for future experimental improvements. To our
knowledge only one explanation has been published to understand
these high temperatures \cite{piilo04}. The argumentation is based
on heating induced by inelastic collisions. However, as the
authors themselves show, this heating mechanism sets in only for
atomic densities much higher than for those measured in typical
MOT ($10^9-10^{11} \mathrm{atoms}/cm^3$). For this reason this
explanation fails to understand the results presented in this
paper and also in all previously cited publications.

Heating mechanisms depending on the density and/or the number of
atoms in the cold MOT cloud are well known even for degenerate
groundstate atoms \cite{drewsen94}. They essentially rely on
reabsorption of scattered photons in the cold MOT \cite{cooper94}.
This situation is encountered as soon as light multiple scattering
sets in, \emph{i.e.} when the light scattering mean free path
$\ell$ is comparable to the MOT size L. Similar multi-atoms and
multi-photons effects certainly also exist with earth-alcaline
MOTs, but they can be made negligible by working at low densities
($n \approx 2.5 \, 10^9 \mathrm{atoms}/cm^3$) and at low optical
thickness ($b=L/\ell \approx 0.3$ at resonance).

Strictly speaking, Doppler theory can also be tested with
degenerate atoms provided a suitable molasses polarization
configuration is chosen \cite{weiss89}. Indeed, for 1D-cooling
with parallel polarizations ($\sigma^+-\sigma^+$ or $\pi-\pi$
configurations), no Sisyphus cooling can occur. However, even in
this case, the authors found unusual high temperatures. They
suggested that this discrepancy was due to molasses intensity
imbalance, leading to a local drift of the atomic average velocity
in the cloud. We agree with this explanation as we will show in
detail in this paper.

We have performed specific temperature measurements on $^{88}Sr$
(Sr in short hereafter) in a 1D-cooling configuration. This is the
ideal situation to test the Doppler theory for two reasons. First,
we induce cooling on a $J_g=0 \to J_e=1$ atomic optical
transition. Second, 2D or 3D Doppler cooling would be more
difficult to analyze quantitatively. Indeed interference between
the six laser MOT beams can induce light shifts modulations,
leading to modifications of the usual friction and diffusion
coefficients \cite{dalibard85}. These spurious effects do not
exist in 1D configuration with mutual orthogonal polarized
counter-propagating laser beams. The new important ingredient here
is that we incorporate in the original Doppler theory the effect
of \emph{transverse spatial intensity fluctuations of the laser
molasses profiles}.

The paper is organized as follows: after some details about the
experimental procedure, we compare velocity dispersion
measurements with the Doppler theory predictions (section
\ref{Experimental Results}). We have found temperatures 10 times
larger than predicted. Indeed, at a laser detuning
$\delta=-\Gamma/2$, we have measured a velocity dispersion
$\sigma_v=0.7\pm0.2 \, m/s$ ($T\approx 5 \, mK$) whereas the
Doppler theory predicts $\sigma_v=0.23 \, m/s$ ($T\approx 0.5 \,
mK$). In section \ref{Cooling with intensity imbalance}, we will
show, on very general arguments, why the molasses transverse
intensity fluctuations heat the cold cloud. We then derive an
analytical model valid in two limiting cases: when the transverse
distance $L_{\perp}$ travelled by an atom during the longitudinal
velocity damping time is much shorter or much longer than the
transverse correlation length $\xi_s$ of the molasses intensity
fluctuations. As expected, these two different limits give rise to
different final velocity distributions. In subsection
\ref{comparison_experimental}, we will compare our results to a
Monte-Carlo simulation. Most of our experimental data correspond
to $L_{\perp} \ll \xi_s$. In this case, the dynamics of the
velocity distribution shows an unusual behavior. The velocity
dispersion $\sigma_v$ is first reduced by the friction force and
then increases after a time scale related to $\xi_s$ (section
\ref{dynamic_cooling}). This specific behavior confirms the
dominant role played by the transverse intensity fluctuations in
1D-cooling with a $J_g=0 \to J_e=1$ transition.

\section{Experimental Results}
\label{Experimental Results}

\subsection{Magneto-optical trap}

The cold strontium cloud is produced in a MOT. The $J_g = 0 \to
J_e = 1$ dipole transition under consideration is the optical
atomic line $^1S_0-{^1P}_1$ at $\lambda = 461 \, nm$. The
excited-state natural linewidth is $\Gamma /2\pi=32 \, MHz$ and
the corresponding saturation intensity is $I_s = 42.5 \, mW/cm^2$.
First an effusive Sr beam is extracted from a $500^{\circ}\,C$
oven. Then a $27 \, cm$ long Zeeman slower reduces the Sr
longitudinal velocity within the velocity capture range of the MOT
($\sim 50 \, m/s$). The Zeeman slower, MOT, and probe laser beams
operate at $461 \, nm$ and are generated from the same
frequency-doubled source detailed in \cite{bruce02}. Briefly, a
single-mode grating stabilized diode laser and a tapered amplifier
are used in a master-slave configuration to produce $500 \, mW$ of
light at $922 \, nm$. This infrared light is then
frequency-doubled in a semi-monolithic standing-wave cavity with
an intra-cavity KNbO$_\mathrm{3}$ nonlinear crystal. The cavity is
resonant for the infrared light while the second harmonic exits
through a dichroic mirror providing $150 \, mW$ of tunable
single-mode light, which is then frequency locked on the $461 nm$
Sr line in a heat pipe. We use acousto-optic modulators for
subsequent amplitude and frequency variations. The MOT is made of
six independent trapping beams. Each beam is carrying an intensity
of $5.2 \, mW/cm^2$ and each beam waist is $8 \, mm$. The trapping
beams are red-detuned by $\delta=-\Gamma$ with respect to the
atomic resonance. Two anti-Helmoltz coils generate a $70 \, G/cm$
magnetic field gradient to trap the atoms. The number of trapped
atoms, as deduced from fluorescence measurements, is $N \approx
2.5 \, 10^6$. The (Gaussian-shaped) cloud dispersion is roughly
$0.6 \, mm$. The velocity dispersion of atoms in the MOT is
typically $\Delta v \sim 1 \, m/s$. Technical details about
temperature measurements are given in the following section.

\subsection{1D-cooling and time-of-flight measurements}

\begin{figure}[]
\includegraphics[scale=0.7]{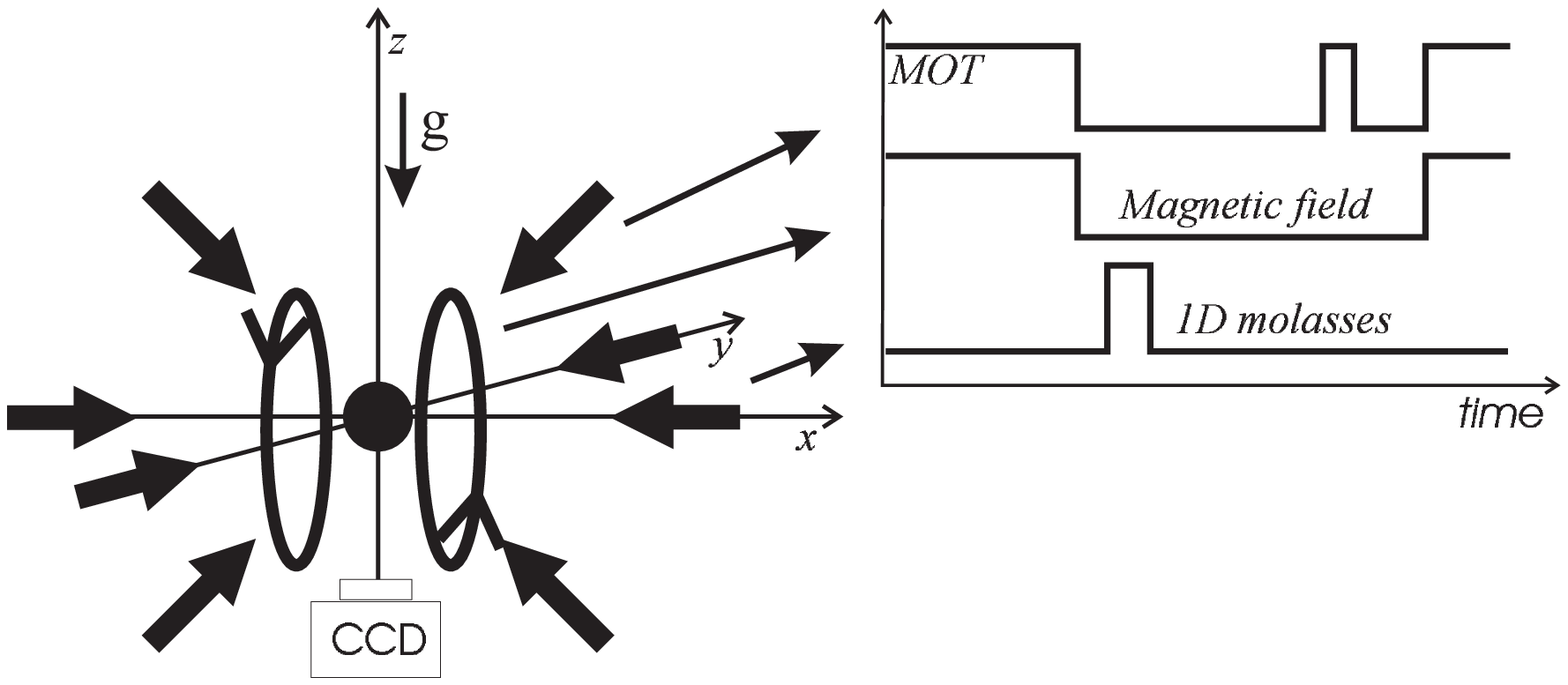}
\caption{Schematic drawing of our set-up and of the time sequence
used in the experiment. The six independent MOT laser beams are
along the $x$ axis and in the vertical plane (yz) at $45^{\circ}$
with respect to the $z$ axis. The one dimensional cooling beams
contrapropagate along the $y$ axis. The recorded CCD images
correspond to the cloud fluorescence signal integrated over $z$.}
\label{set_up}
\end{figure}

In addition to the previously discussed experimental set-up (MOT,
Zeeman slower), two additional contra-propagating laser beams are
used to perform a 1D molasses (see figure \ref{set_up}). These 1D
cooling beams are located in the horizontal plane. Some
quarter-wave and half-wave plates are used to fix the relative
polarizations of the two contra-propagating laser beams. An
acousto-optical modulator in a double-pass configuration is used
to adjust the laser frequency from resonance down to $-\Gamma$.
The same acousto-optical modulator is used to control the laser
intensity up to $I_s$. The beam waist is $3.5 \, mm$ at the cold
cloud position, thus much bigger than the MOT size.

The time sequence of the experiment is generated by a personal
computer with digital output ports dedicated to this task. An
internal clock updates and synchronizes the digital output ports
every $10 \,\mu s$. This elementary time step is short enough for
our purposes. The digital output ports are connected to switches
which turn on and off the lasers beams (rising and falling times
shorter than $1 \,\mu s$), the magnetic field (rising time of few
$ms$ and falling time of $100 \,\mu s$) and the CCD camera chopper
(opening and closing times of $200 \,\mu s$). The time sequence is
designed as follows: first the MOT is operated during about $20
\,ms$. Then the MOT lasers and magnetic gradient are switched off.
The 1D-molasses laser beams are then switched on during $500 \,\mu
s$. This cooling time is appropriately chosen : atoms reach the
cooling steady-state regime while, at the same time, the expanding
atomic cloud remains smaller than the cooling beams size. After
the 1D-cooling sequence, the temperature of the cold cloud is
extracted from a time-of-flight (TOF) measurement technique. For
this purpose, all the laser beams are switched off and the cold
cloud expands ballistically in the dark. The duration of the dark
period is varied from a few $100 \, \mu s$ up to $1.5 \,ms$. Then
a fluorescence image of the expanding cloud is recorded on a CCD
camera by briefly switching on the MOT beams during $20 \,\mu s$.
The whole time sequence is then repeated as long as necessary to
obtain a good signal-to-noise ratio (see figure \ref{set_up}).

\begin{figure}[]
\includegraphics[scale=0.5]{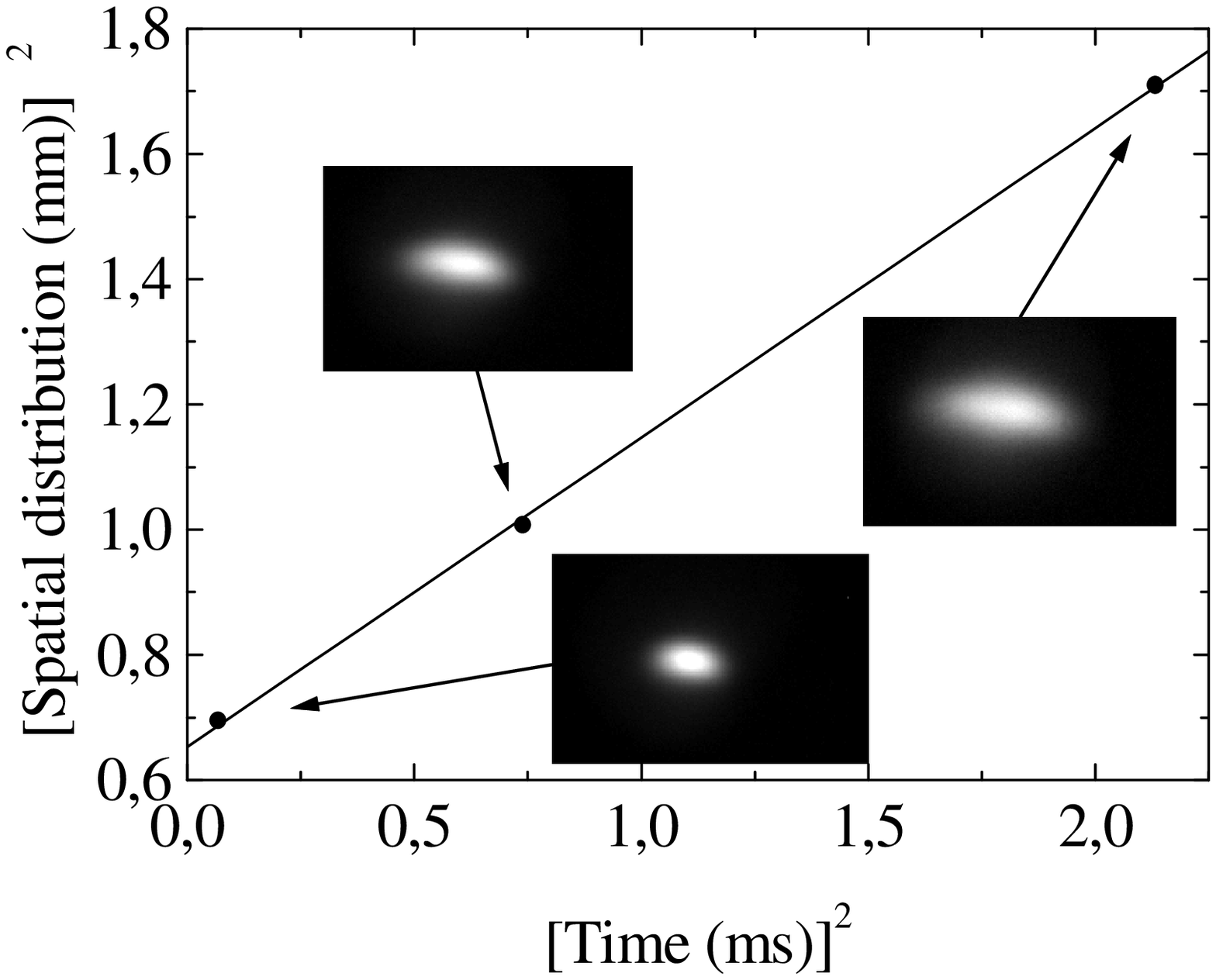}
\caption{After the 1D-cooling sequence along axis $y$, the spatial
distribution in the horizontal plane is collected at three
different ballistic expansion times by the time-of-flight
technique. From these images, we extract the velocity dispersion
$\sigma_v$ in the cloud (see text).} \label{TOF}
\end{figure}

Figure \ref{TOF} shows three images of the cloud after different
ballistic expansion times. The elliptical shape of the cloud is a
signature of the 1D-cooling sequence. Indeed the cooling axis is
precisely the small axis of the ellipse. This dimension will be
thereafter referenced as the \emph{longitudinal} axis. In the two
other dimensions, thereafter referenced as the \emph{transverse}
axes, the cloud is heated by random spontaneous emission. From
these images, we extract the dispersion of the longitudinal
\emph{spatial} distribution of atoms in the cloud, integrated over
the transverse directions. Because of the finite size of the
image, we cannot have access to the long tail of the spatial
distribution. Hence, we truncate all recorded distributions at
$2\%$ of their maximum value. Assuming the initial atomic
positions and velocities in the MOT to be uncorrelated and
centered, the ballistic time evolution of the spatial dispersion
is simply $\sigma_x^2(t)=\sigma_x^2(0)+\sigma_v^2\,t^2$. The
velocity dispersion $\sigma_v = \sqrt{\langle v^2\rangle}$ is then
easily extracted from the experimental data. To cross-check these
measurements, we have also used a different measurement method,
namely a spectroscopic technique. It consists in probing the cold
cloud with an ultra-stable laser beam tuned on resonance with the
$^1S_0 - {^3P}_1$ Sr transition line (note that this is also a
$0\to1$ transition). Because this line is spin-forbidden, its
frequency width is very small ($7.5 \,kHz$) and is thus
Doppler-broadened in the MOT \cite{paper_red}. Hence, we have a
direct access to the velocity distribution by measuring the
Doppler-induced spectral width. We have checked that these two
different techniques give the same results, thereby confirming the
validity of the TOF measurements, easier to handle, and for this
reason routinely used in this experiment for temperature
measurements.

\subsection{Doppler theory and results}

Doppler theory is based on absorption-fluorescence cycles which
induce a cooling mechanism competing with a heating mechanism.
Cooling is due to a mean friction force $F = -m\gamma v$ which
damps the velocity and heating is due to a Langevin fluctuating
force $F_{\nu}$ (photon noise) giving rise to diffusion with
constant $D$ \emph{in velocity space}. At equilibrium these
processes exactly balance (fluctuation-dissipation theorem) and
the Doppler temperature is found to be $k_BT_D = m \, \sigma_D^2 =
m\,D/\gamma$. All experimental data presented in this paper have
been obtained in the $lin\|lin$ polarization configuration. In
this case, analytical expressions for the damping coefficient
$\gamma$ can be found in \cite{gordon80} and in \cite{lett} for
the diffusion constant $D$. The analytical expression of the
Doppler variance at low laser intensities $I$ and \emph{negative}
detuning $\delta$ is then
\begin{equation}\sigma_D^2 = \frac{D}{\gamma}\approx\frac{7}{20}\;\frac{\hbar\Gamma}{m}\;\frac{1+\Delta^2}{2|\Delta|}\;\left(
1+\frac{2}{7}\, \frac{11+\Delta^2}{(1+\Delta^2)^2}\; s_0\right) =
\sigma_0^2(\Delta)\;(1+2\beta (\Delta) \, s_0) \end{equation} Here
$\Delta=2\delta /\Gamma$ is the laser detuning in units of
$\Gamma/2$ and $s_0=I/I_s$ is the on-resonance saturation
parameter. As could have been guessed, the Doppler temperature is
related to the only energy scale of the problem, namely the
excited-state energy width $\hbar\Gamma$. We thus see that, at low
intensities,
 $\sigma_D \approx \sigma_0(\Delta) \,(1+\beta(\Delta) s_0)$ grows linearly in $s_0$ with slope $\beta (\Delta) \sigma_0(\Delta)$
 from $\sigma_0(\Delta)$. Minimization of $\sigma_0(\Delta)$ is achieved for $\Delta =-1$ and gives $\beta = 3/7$, $\sigma_0 =
\sqrt{7\hbar\Gamma/20m} \approx 0.23 \,m/s$.

In the strict 1D Doppler theory, the final velocity distribution
and the corresponding temperature should depend on the
polarization configuration. For example, in the $lin\|lin$
polarization channel, photon redistribution processes between the
two contra-propagating laser fields can occur, whereas they are
forbidden in the $lin\perp lin$ polarization channel. Hence, the
polarization configuration affects both friction and diffusion. We
have tested different polarization configurations ($lin\|lin$,
$lin\perp lin$ and $\sigma^+-\sigma^-$) but no significative
temperature modifications have been found in the parameters range
used in this experiment (see below).

\begin{figure}[]
\includegraphics[scale=0.5]{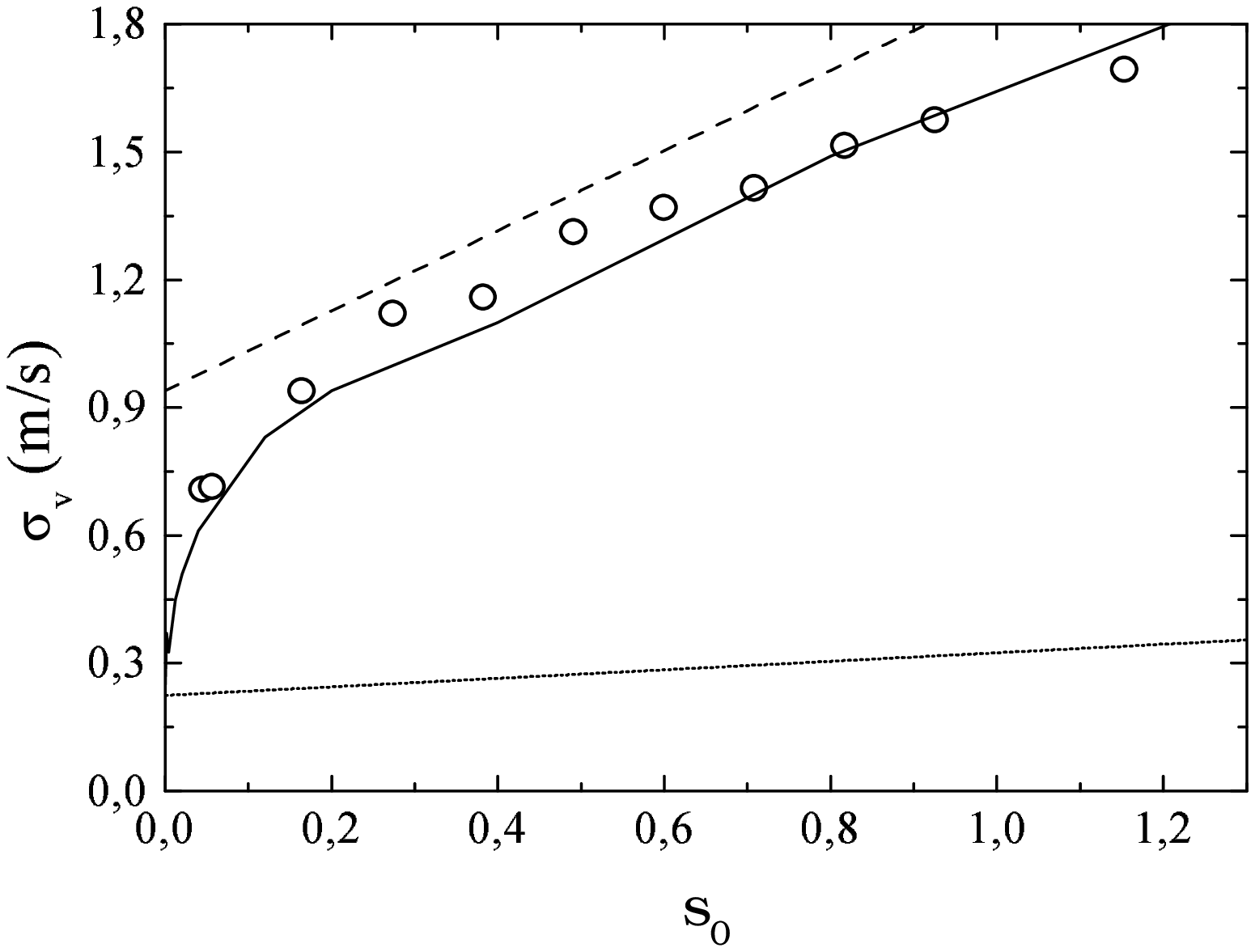}
\caption{Velocity dispersion $\sigma_v$ as a function of the mean
saturation parameter $s_0=I/I_s$ at  $\Delta =2\delta /\Gamma=-1$.
The experimental data (circles) are compared to the Doppler
prediction (dotted line) and to the Monte-Carlo simulation (solid
line) at $r_s=9\%$ and $\xi_s=60 \,\mu m$. Whereas the Doppler
theory is completely off, very good agreement is found with our
theoretical model. The dashed line prediction corresponds to
equation (\ref{long c}) (see text).} \label{intensity}
\end{figure}

Figure \ref{delta} shows the experimental variation of the
velocity dispersion $\sigma_v$ as a function of $\Delta$ at low
intensity ($s_0=0.08$). The global behavior is the same as the one
predicted by the Doppler theory but systematically higher. For
example, at $\Delta =-1$, we get $\sigma_v=0.7\pm0.2 \,m/s$
($T\approx5 \,mK$) whereas the Doppler theory predicts $\sigma_v
\approx \sigma_0\approx 0.23 \,m/s$ ($T\approx0.5 \,mK$). The
mismatch is even more pronounced at small detuning and tends to be
reduced at high detuning. Figure \ref{intensity}, obtained at
$\Delta = -1$, shows a strong linear increase of $\sigma_v$ as a
function of $s_0$. The measured slope is $0.9 \,m/s$ whereas the
Doppler theory only predicts $0.1 \,m/s$.

The on-resonance optical thickness corresponding to the data in
figures \ref{delta} and \ref{intensity} is typically $b\approx
0.3$ when the experimental sequence starts. At the end of the
sequence, because the atomic cloud expands, $b\approx 0.15$. These
optical thicknesses are not sufficiently low to discard multiple
scattering effects (this would require $b \ll 1$). The role of
multiple scattering of light is two-fold. First, it induces photon
reabsorption leading to an average repulsion force and heating.
However at $b\approx 0.3$, heating, as observed in
\cite{drewsen94}, can here be neglected. Second, it implies beam
attenuation (Lambert-Beer law). This means that the average
optical force is weaker for atoms located deep in the cloud and we
get, as a net effect, an average compression force. In a MOT,
these compression and repulsion forces are equilibrated by the
trapping force \cite{sesko91}. This is not the case in optical
molasses. In a 1D-molasses, the compression force is expected to
dominate over the repulsion force induced by photon reabsorption,
at least at moderate optical thicknesses. A quantitative estimate
of multiple scattering effects is thus not easy. For example, in
our case, the Lambert-Beer law predicts a maximum relative
intensity imbalance of approximately $23\%$ at low intensity,
which is not negligible. However our experimental results strongly
suggests that multiple scattering velocity inhomogeneous
broadening is small. Indeed, as $s_0$ is increased, the optical
thickness decreases since the scattering cross-section is reduced.
The compression force is thus decreased and cannot explain the
strong increase of $\sigma_v$ {\it vs} $s_0$ evidenced in figure
\ref{intensity}. Furthermore, multiple scattering induces some
correlations between position and velocity which should alter the
cloud ballistic expansion. This has not been observed in figure
\ref{TOF}. As an ultimate test, we have change the number of atoms
in the MOT by a factor $3$ without detecting any modification of
$\sigma_v$.

\section{Cooling with intensity imbalance}
\label{Cooling with intensity imbalance}

We analyze in this section how \emph{spatial stationary}
transverse intensity fluctuations can modify the Doppler cooling
predictions in a quantitative way. Starting from the analytical
expression of the average force in 1D molasses, we take into
account these spatial intensity fluctuations and we derive
analytical results in two limiting cases, namely when cooling is
achieved before intensity fluctuates and in the opposite case.
This analytical model neglects the photon noise encapsulated in
the Langevin force $F_{\nu}$ leading to a nonzero Doppler
temperature. However, for a quantitative comparison with
experimental data, we have developed a Monte-Carlo simulation
which fully takes into account all fluctuating mechanisms
(molasses intensity fluctuations \emph{and} photon noise). These
results will be detailed in subsections \ref{Monte-Carlo} and
\ref{comparison_experimental}.

\subsection{Origin of spatial intensity imbalance}
\label{Orign_imba} 
\begin{figure}[]
\includegraphics[scale=0.7]{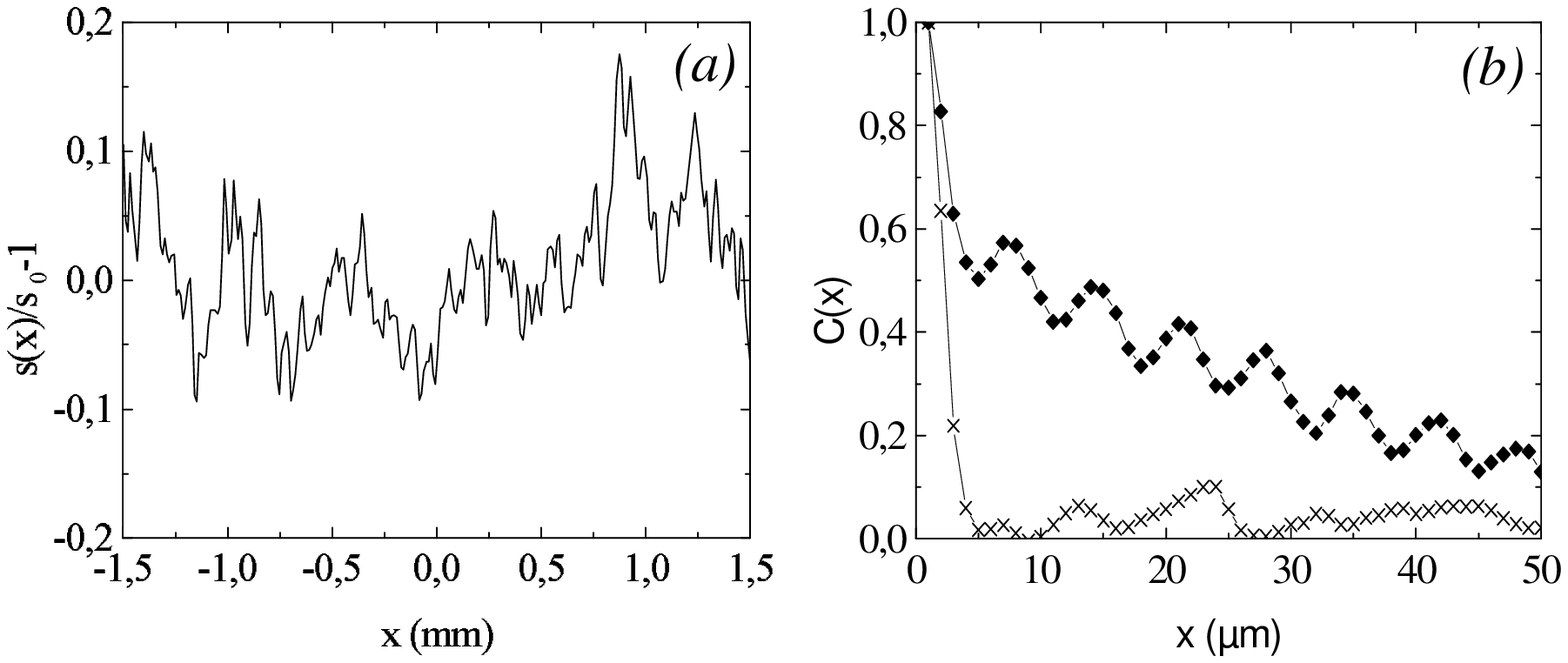}
\caption{Figure (a) shows typical relative intensity fluctuations
of a laser beam once the smooth Gaussian profile is removed. The
typical intensity dispersion is roughly $10\%$ of the total
signal. Figure (b) shows the spatial fluctuations correlation
function $\mathcal{C}(x)$ \emph{vs} a transverse coordinate. The
transverse distance at which this correlation function vanishes
defines the correlation length $\xi_s$. Experimentally we found
$\xi_s \simeq 30 \,\mu m$ (diamonds). The oscillation on top of
the decay is due to the interference generated by the front window
of the CCD camera. The fast decrease observed at short distances
is also present when the laser is off (crosses). This fast
decrease is thus due to the uncorrelated noise on the CCD camera
and corresponds to the pixel size.} \label{putre}
\end{figure}

In the standard Doppler theory, the molasses
beams are described as perfect plane waves. In real experiments
however, the beams have a Gaussian-shaped transverse profile.
Usually one can ignore the transverse dimensions of the beams
because they are generally much larger than the MOT size.
Nevertheless, the transverse intensity profiles are not
defect-free ideal Gaussian profiles. Indeed, even starting from
diffraction-limited laser beams, imperfections of optics elements
(dust, aberrations, \emph{etc}) induce scattering which generates
an intensity speckle. Of course, in well-controlled experiments,
this speckle pattern remains relatively small compared to the
average beam intensity, but those two fields add coherently. In
our experiment, we have measured the laser beam spatial transverse
fluctuations by placing a CCD camera at the approximative position
of the MOT. Subtracting the ideal Gaussian profile, we have
computed the intensity fluctuations histogram (see figure
\ref{putre}). We found a Gaussian histogram with a relative
standard deviation (with respect to the average intensity) in the
range $10-20\%$.

The 1D-molasses is created by two contra-propagating beams issued
from the same source. However, the speckle intensity generated by
the optics imperfections in each arm are independent. It is thus
reasonable to consider, at each transverse position in the beam
profile, the intensities $I_i \,(i=1,2)$ of the two molasses beams
to be random \emph{independent} Gaussian variables. Noting by
$s_i$ the corresponding on-resonance saturation parameters, we
assume the probability distribution $\mathcal{P}(s_i)$ to be the
same for the two beams. Hence the (common) first two moments, at
each transverse position, are $\langle s_i\rangle = s_0$ and
$\sigma_s^2 = \langle (s_i-s_0)^2 \rangle$ $(i=1,2)$. In the
following we will characterize the molasses intensity fluctuations
by the ratio $r_s = \sigma_s/2s_0$. From experimental data, we
have $r_s$ in the range $5-10\%$. Another important feature is the
transverse spatial correlation function (assumed to be the same
for each beam), namely $\mathcal{C}(\textbf{r})=\langle
s_i(\textbf{r}')\,s_i(\textbf{r}'+\textbf{r})\rangle \, (i=1,2)$.
The characteristic decay length of $\mathcal{C}(\textbf{r})$
defines the transverse correlation length $\xi_s$. Strictly
speaking, because of diffraction effects, $\xi_s$ cannot be zero.
We have experimentally found $\xi_s$ to be of the order of few
tenth of $\mu m$ ($\xi_s=30 \mu m$ for the example shows in figure
\ref{putre}).

\subsection{Analytical model}
\label{analytic_model}

For a $0 \to 1$ transition in the $lin\|lin$ configuration, atoms
are modelled by a two-level system involving the same Zeeman
states for each molasses beam. The two beams have the \emph{same}
red-detuned frequency ($\delta<0$), \emph{opposite} wave-vector
$\textbf{k}$ but locally \emph{different} saturations which
fluctuate independently across the transverse profile of the beams
but with the \emph{same} correlation length $\xi_s$. The mean
local radiation pressure force experienced by a single atom, with
longitudinal velocity $v$, takes the simple following form:
\begin{eqnarray}
F&=&m\,\frac{v_R\Gamma}{2}\left\lbrack\frac{s_1}{1+s_1+s_2+(\Delta-2kv/\Gamma)^2}-
\frac{s_2}{1+s_1+s_2+(\Delta+2kv/\Gamma)^2}\right\rbrack
\\
&\approx&
m\,\frac{v_R\Gamma}{2}\left\lbrack\frac{s_1}{1+2s_0+(\Delta-2kv/\Gamma)^2}-
\frac{s_2}{1+2s_0+(\Delta+2kv/\Gamma)^2}\right\rbrack
\label{force_exact}\end{eqnarray} where $v_R=\hbar k/m$ is the
recoil velocity ($\approx 6 \, mm/s$ for Sr). We have here assumed
that the mechanical actions of the two laser beams can be added
independently, only considering the total saturation of the
optical transition to be $(s_1+s_2)$. We have furthermore replaced
($s_1+s_2$) by $2s_0$ in the denominator. A more careful analysis,
along the lines given below, shows that this is indeed valid with
our experimental parameters.

Near the steady state, the Doppler broadening become negligible,
$kv\ll(\Gamma,\delta)$ and the force can be safely approximated
by:
\begin{eqnarray}
F &\approx& -m (s_1+s_2)\gamma_v\,v+m (s_1-s_2)\, a \label{force} \\
&\approx& -2ms_0\gamma_v\,v+m (s_1-s_2)\, a
\label{forceapprox}\end{eqnarray} Defining the recoil angular
frequency $\omega_R = 2\pi\,\nu_R = \hbar k^2/2 m$ ($\nu_R \approx
10.6 \,kHz$ for Sr), we have
\begin{equation}
\label{coeff}
\gamma_v=\omega_R\,\frac{4|\Delta|}{(1+\Delta^2+2s_0)^2}\quad\textrm{and}\quad
a=\frac{v_R\Gamma}{2}\, \frac{1}{1+\Delta^2+2s_0} \end{equation}
The friction force is given by the term proportional to $v$ and
gives rise to a mean damping time:
\begin{equation}
\label{damping} \tau_v=(2s_0\gamma_v)^{-1}
\end{equation}

Consider now an atom moving across the molasses beams with
transverse velocity $v_{\perp}$. This velocity is unaffected by
the longitudinal cooling. So, as the atom flies across the beam,
it may experience \emph{induced temporal} molasses intensity
fluctuations. The induced \emph{correlation time} of these
fluctuations is simply
\begin{equation}\tau_s=\xi_s/v_{\perp} \label{TfL}\end{equation}
We have thus two competing dynamical processes, the cooling one
with characteristic time $\tau_v$ and the intensity random
(induced) temporal variations with correlation time $\tau_s$. The
global velocity dynamics can be easily analyzed when these time
scales are well separated.

\subsubsection{The regime $\tau_v \ll \tau_s$}

For atoms fulfilling this condition, the damping process is
completed before intensities fluctuate. Then the constant force
and the friction coefficient in (\ref{force}) can be considered as
time-independent. Atoms thus behave as if $v_{\perp}$ was zero.
However, because $\xi_s$ is much shorter than the cloud size, each
atom experiences randomly fixed distributed intensity imbalances
and, in turn, a random constant force. The random stationary
velocity is $v_{\infty}=a\tau_v(s_1-s_2)$. It is
Gaussian-distributed with zero mean and dispersion
\begin{equation} \label{long c}
 \sigma_{\infty}=\sqrt{2}a\tau_v\,\sigma_s =
\sqrt{2} \; \frac {\Gamma}{k} \; \frac {1+|\Delta |^2 +
2s_0}{4|\Delta |} \; r_s
\end{equation} To find the
final velocity distribution, one has to incorporate the photon
noise force $F_{\nu}$. As the photon noise and intensity
fluctuations are independent, the total velocity variance is the
sum of the Doppler and intensity variances
$\sigma_v^2=\sigma_D^2+\sigma_{\infty}^2$. In this regime,
intensity fluctuations thus give rise to an extra-heating
mechanism inducing an inhomogeneous broadening of the Doppler
velocity distribution. At $\Delta =-1$, $s_0$ small and $r_s =
9\%$, one gets $\sigma_{\infty} \approx 0.94 \, m/s$ and $\sigma_v
\approx 0.97 \, m/s$.

\subsubsection{The regime $\tau_v \gg \tau_s$}

In this regime, the molasses intensities fluctuate wildly before
the atom reaches its stationary state. Thus they cannot be treated
as time-independent quantities. They can however be treated as
independent \emph{Markovian} (short memory) processes. From the
expression (\ref{force}) of the total force, we derive the
following master equation for the velocity distribution
$\mathrm{P}_t(v)$ \cite{van_kampen} (details will be given
elsewhere):
\begin{equation}
\mathrm{P}(v,t+\tau_s)=\int\dd s_1\,\dd s_2\,\dd
v'\,\mathcal{P}(s_1)\, \mathcal{P}(s_2)\,
\mathrm{P}(v',t)\,\delta\left(v-v'-\frac{F}{m}\tau_s\right)\
 \label{master}\end{equation} In this equation, at each time step $\tau_s$, the saturations $s_1$
and $s_2$ take on new values uncorrelated with the previous ones.

One should note that we have kept here both the \emph{additive}
noise, given by the $(s_1-s_2)$ term, and the
\emph{multiplicative} noise, given by the $(s_1+s_2)v$ term. This
last term had been replaced by $2s_0v$ in our previous analysis
(expression (\ref{forceapprox}) of the force). In the limit of
small velocity changes at the elementary time scale $\tau_s$, one
can derive from (\ref{master}) a Fokker-Planck type equation (see
Appendix \ref{DerivMasterEq}):
\begin{equation}\frac{\partial \mathrm{P}(v,t)}{\partial t}
=\frac{1}{\tau_v}\frac{\partial}{\partial
v}(v\mathrm{P}(v,t))+\frac{\partial^2}{\partial
v^2}(D(v)\mathrm{P}(v,t)) \label{FockkerP}\end{equation} Defining
$D_{\infty} = \sigma_{\infty}^2/2\tau_v$, the velocity-dependent
diffusion constant has the following expression:
\begin{equation}D(v)= D_{\infty} \; \big{[}1+(1+2r_s^2) \left\lbrack v/\sigma_{\infty}
\right\rbrack ^2 \big{]} \; \frac{\tau_s}{\tau_v} \approx
D_{\infty} \; \big{[}1+(v/\sigma_{\infty}) ^2 \big{]} \;
\frac{\tau_s}{\tau_v}
\end{equation} since generally $2r_s^2 \ll 1$. This velocity-dependence originates from the
multiplicative noise and leads to abnormal diffusion. The
stationary velocity distribution $\mathrm{P}_0$ is then easily
found to take the following form ($\mathcal{N}$ is a normalization
constant):
\begin{equation}
\mathrm{P}_0(v)\approx \mathcal{N}\,[ 1+(v/\sigma_{\infty})^2
]^{-(1+\eta)}
\end{equation} where $\eta = \tau_v/\tau_s \gg 1$. The full width at
half maximum of this distribution is
\begin{equation}\label{short_c}
\Gamma_v = 2\sqrt{\ln{2}} \, \sqrt{\frac{\tau_s}{\tau_v}} \;
\sigma_{\infty} \: \ll \: \sigma_{\infty}
\end{equation} As easily checked $D(v) \approx D_{\infty} \; \tau_s/\tau_v$ for $v \sim
\Gamma_v$. Hence the impact of abnormal diffusion is in fact
negligible when $\tau_v \gg \tau_s$ and the velocity distribution
is very close to a Gaussian with velocity dispersion
\begin{equation}
\label{short_c}
\sigma_v\sim\sqrt{\frac{\tau_s}{\tau_v}}\,\sigma_{\infty} \ll
\sigma_{\infty}
\end{equation}
which is much narrower than the velocity dispersion found in the
long correlation limit (\ref{long c}) by the factor
$\sqrt{\tau_s/\tau_v}$.\\

\begin{figure}[]
\includegraphics[scale=0.5]{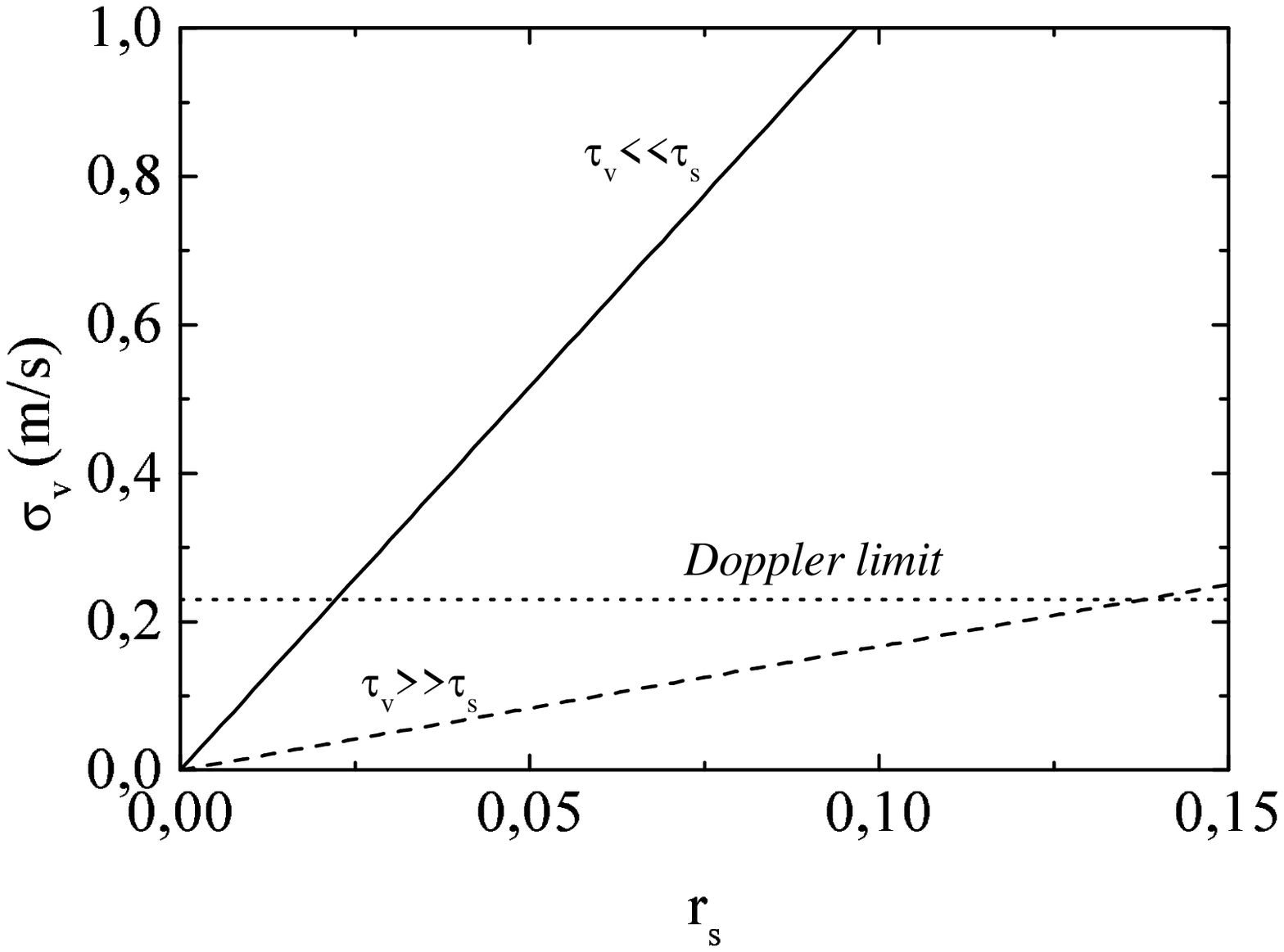}
\caption{velocity dispersion $\sigma_v$ for the long (solid line)
and short (dashed line) correlation time limit as a function of
the relative fluctuation of the intensity $r_s=\sigma_s/2s_0$. In
the short correlation time limit $\tau_s=30 \mu s$. The dotted
line corresponds to the Doppler limit at  low saturation and
$\Delta=-1$ (see text).} \label{analytic}
\end{figure}

Figure \ref{analytic} shows $\sigma_v$ as a function of $r_s$ in
the long and short correlation time limits ($\Delta=-1$, $s_0\ll
1$). When $r_s \to 0$, $\sigma_v$ goes to zero in both limits.
This does not mean that the temperature goes to zero in
experiments. Indeed, in the previous discussion we have discarded
the photon noise at the physical origin of the bare Doppler
theory. Its contribution at low saturation is given by the dotted
line in figure \ref{analytic}. We can see that the noise induced
by long correlation times starts to dominate at $r_s
> 2.5 \%$, a fairly small value. Hence, to achieve cooling up to the
Doppler limit at $\Delta=-1$, one has to minimize the ratio
$\tau_c/\tau_v$ for a fixed value of $r_s$. This can be done in
different ways. First, by decreasing the laser intensity to
increase $\tau_v$. Second, by decreasing $\tau_s$. As shown by
(\ref{TfL}), this can be achieved either by avoiding large
transverse intensity defects in the laser profiles to decrease the
correlation length $\xi_s$ or by increasing $v_{\perp}$. This is
indeed was has been observed in atomic beam experiment
\cite{witte92} where temperatures very close to the Doppler limit
had been found along one \emph{transverse} dimension.

In a MOT, if the fluctuations of the transverse intensities are
not the same for all three dimensions, the temperature is expected
to be anisotropic. Moreover coupling mechanisms occur between the
cooling dimensions. Thus high velocity dispersion along one
dimension tends to reduce the temperature in the orthogonal plane.
In our 1D-molasses, the transverse velocity is fixed by the
initial MOT sequence and $\tau_s/\tau_v$ is usually bigger than 1.
As we will see in Section III.D., this means that transverse
intensity fluctuations are the major heating mechanism.

\subsection{Monte-Carlo simulations}
\label{Monte-Carlo}

In order to quantitatively test our theory and not relying on
questionable approximations, we have developed a Monte-Carlo (MC)
simulation. This MC simulation fully takes into account the photon
noise leading to the Doppler cooling limit, saturation of the
transitions and the transverse intensity fluctuations discussed so
far, {\it i.e.} with \emph{arbitrary} correlation time $\tau_s$
and saturation dispersion $\sigma_s$.

The shortest time scale in the MC calculation is given by the
excited-state lifetime $\tau_{e}=1/\Gamma$. To mimic the
transverse flight of atoms, the molasses saturation fluctuations
$\delta s_i = (s_i-s_0)$ ($i=1,2$) become time-dependent
parameters and evolve according to the damped random path discrete
equation ($n$ labels the number of time steps $\tau_e$)
\begin{equation}
\label{random_I} \delta s_i(n+1)=(1-\rho)\,\delta s_i(n)+ R_n
\end{equation}
with $\rho \ll 1$. The first term in this equation is the friction
term, relaxing saturation to its stationary value $s_0$, while the
last term is a random variable, with zero mean value, uniformly
distributed over the range $[-\epsilon /2,\epsilon /2]$. A simple
calculation shows that $\langle R_n^2 \rangle = \epsilon^2/12$. We
further assume that the $R_n$'s are decorrelated. The continuous
limit of (\ref{random_I}) is
\begin{equation}
\label{cont rand} \frac{d\,\delta s_i}{dt} +
\frac{\rho}{\tau_e}\,\delta s_i = R(t)
\end{equation}
As the correlation time $\tau_s$ should correspond to the damping
time of this equation, we see that $\rho=\tau_e/\tau_s$. Thus
fixing $\tau_s$ fixes $\rho$ in (\ref{random_I}). The last term of
(\ref{cont rand}) is a $\delta$-correlated Langevin term with zero
mean. Its time-correlation function is $\langle R(t') R(t) \rangle
= 2\mathcal{D} \, \delta (t-t')$ where it is easily shown that
$\mathcal{D}=\epsilon^2/24\tau_e$. The fluctuation-dissipation
theorem then dictates $\sigma_s^2 = \mathcal{D}\tau_s$, leading to
\begin{equation} \epsilon = 2\sqrt{6\rho} \, \sigma_s \end{equation}
This last result can be found more elegantly by squaring
(\ref{random_I}) and averaging over the probability distribution
of $R_n$. Hence experimental determination (or convenient choice)
of the macroscopic ingredients $\tau_s$ and $\sigma_s$ fixes in
principle the microscopic ingredients $\epsilon$ and $\rho$ in
(\ref{random_I}).

\begin{figure}[]
\includegraphics[scale=0.5]{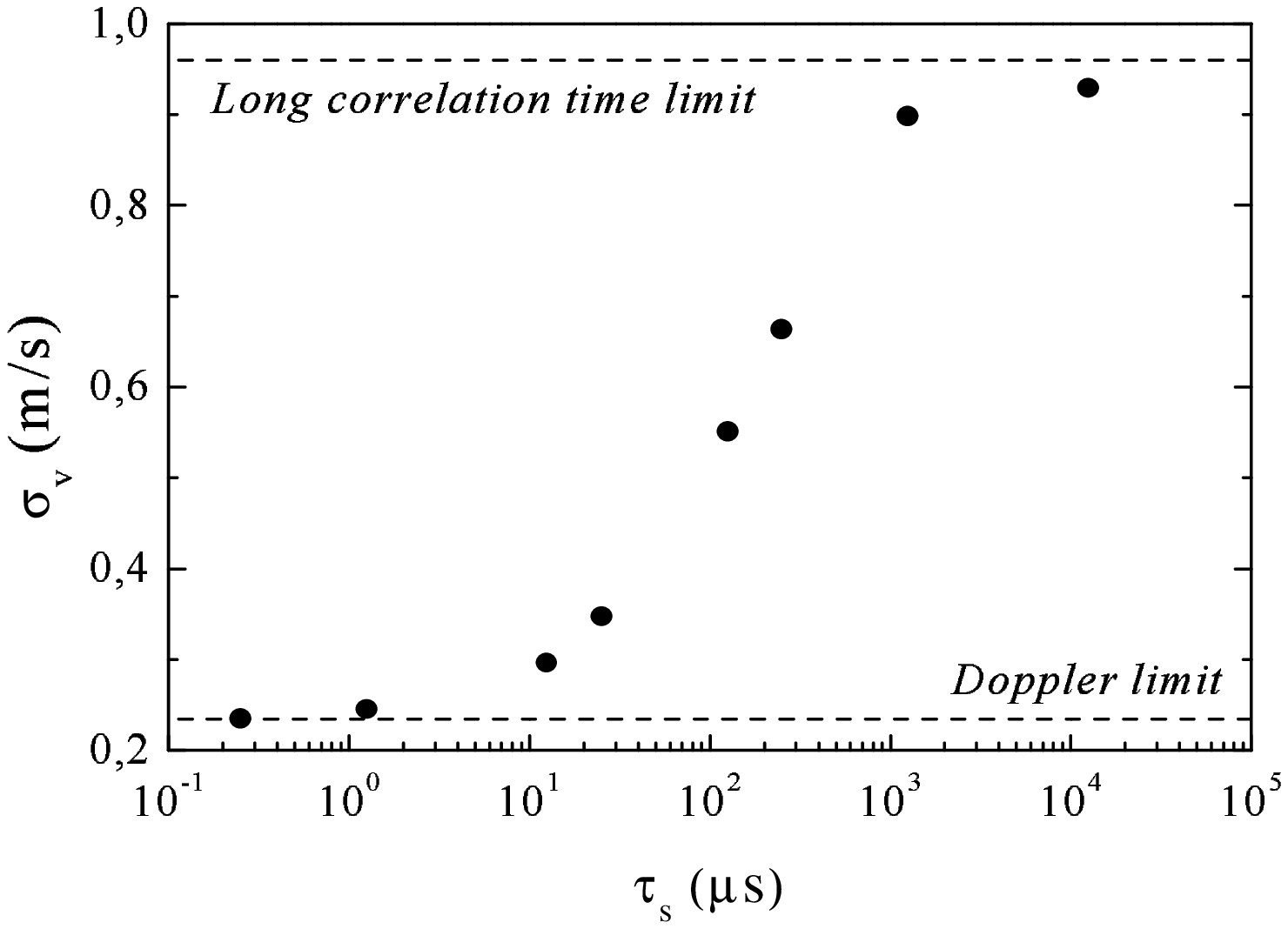}
\caption{Final velocity dispersion $\sigma_v$ as a function of the
correlation time $\tau_s$ (full circles). The transverse and
longitudinal velocity dispersions are both equal to $\Delta v=0.8
\,m/s$. The mean saturation of each beam is $s_0=0.04$ and the
saturation fluctuation parameter is $r_s=9\%$. The upper dashed
line represents the long correlation time limit while the lower
one represents the Doppler-cooling limit (see text).} \label{Coor}
\end{figure}

Figure \ref{Coor} shows the obtained final velocity dispersion
$\sigma_v$ as a function of $\tau_s/\tau_v$ for $s_0 =0.04$ and
$r_s=9\%$. The velocity damping time, calculated with
(\ref{damping}), is $\tau_v=200 \,\mu s$. As expected, $\sigma_v$
is higher for long correlation times $\tau_s \gg \tau_v$. When
$\tau_s \sim \tau_v$, $\sigma_v$ is very sensitive to $\tau_s$. At
lower values $\tau_s \ll \tau_v$, $\sigma_v$ is minimum and, for
these parameters, reaches the Doppler limit. The final
distributions (not shown here) are quasi-Gaussian, even at small
$\tau_s$.

\subsection{Quantitative comparison with experimental data}
\label{comparison_experimental}

We have showed in section II.C that the measured velocity
dispersion were always larger than the Doppler theory predictions.
We now compare our experimental data with the results of our
previous MC simulation. In order to stick as close as possible to
the experiments, we also take into account the transverse velocity
distribution (a centered Gaussian with dispersion $\Delta
v_{\perp} \approx 0.8 \, m/s$) in the MC simulation.

In figure \ref{delta} the solid curve corresponds to the MC
simulation performed at $r_s=9\%$ and $\xi_s=60\,\mu m$. These
quantities are fit-parameters in the MC simulation but remain in
the range of the measured ones (see \ref{Orign_imba}). As one can
see, the agreement with experimental points is now very good,
providing a clear understanding of the physics at work in the
experiment. Coming back to (\ref{coeff}), close to resonance, the
friction term decreases while the intensity imbalance term
increases. Hence, the mismatch between experiment and Doppler
theory is maximal.

Again, an excellent agreement between MC simulation (solid curve)
and data is found in figure \ref{intensity}. To properly
understand these results, one has to remind that the cross-over
region between the long- and short-correlation time limits is
given by $\tau_s=\tau_v$, or equivalently by $s_0=v_{\perp}/
2\gamma_v \xi_s \sim 0.1$. TOF measurements before the molasses
sequence have shown that $v_{\perp} \approx 0.8 \, m/s$ whereas
$\xi_s$ is the chosen MC parameter. This is exactly what is seen
in figure \ref{intensity} and the Doppler theory is recovered when
$s_0 \to 0$. However, this short-correlation time limit
($\tau_s\ll \tau_v$) was not experimentally accessible because the
1D-cooling duration sequence was not long enough to reach the
steady state. This is why most of the data points correspond to
the long-correlation time limit ($\tau_s \gg \tau_v$) and can be
compared to (\ref{long c}). This is done in figure \ref{intensity}
where the dashed line corresponds to the prediction given by
(\ref{long c}). The general behavior is correct, indicating that
saturation of the atomic transition indeed plays a significant
role. However the prediction (\ref{long c}) is a little bit too
large. This is not surprising since, for the explored range of
parameters, some atoms will always have sufficiently high
transverse velocities to fulfill the short correlation time
criterion, both in the experiment and in the MC simulation. As a
consequence, the velocity dispersion $\sigma_v$ will be reduced.
In other words, (\ref{long c}) corresponds to a zero-transverse
velocity case giving over-estimated predictions.

\begin{figure}[]
\includegraphics[scale=0.5]{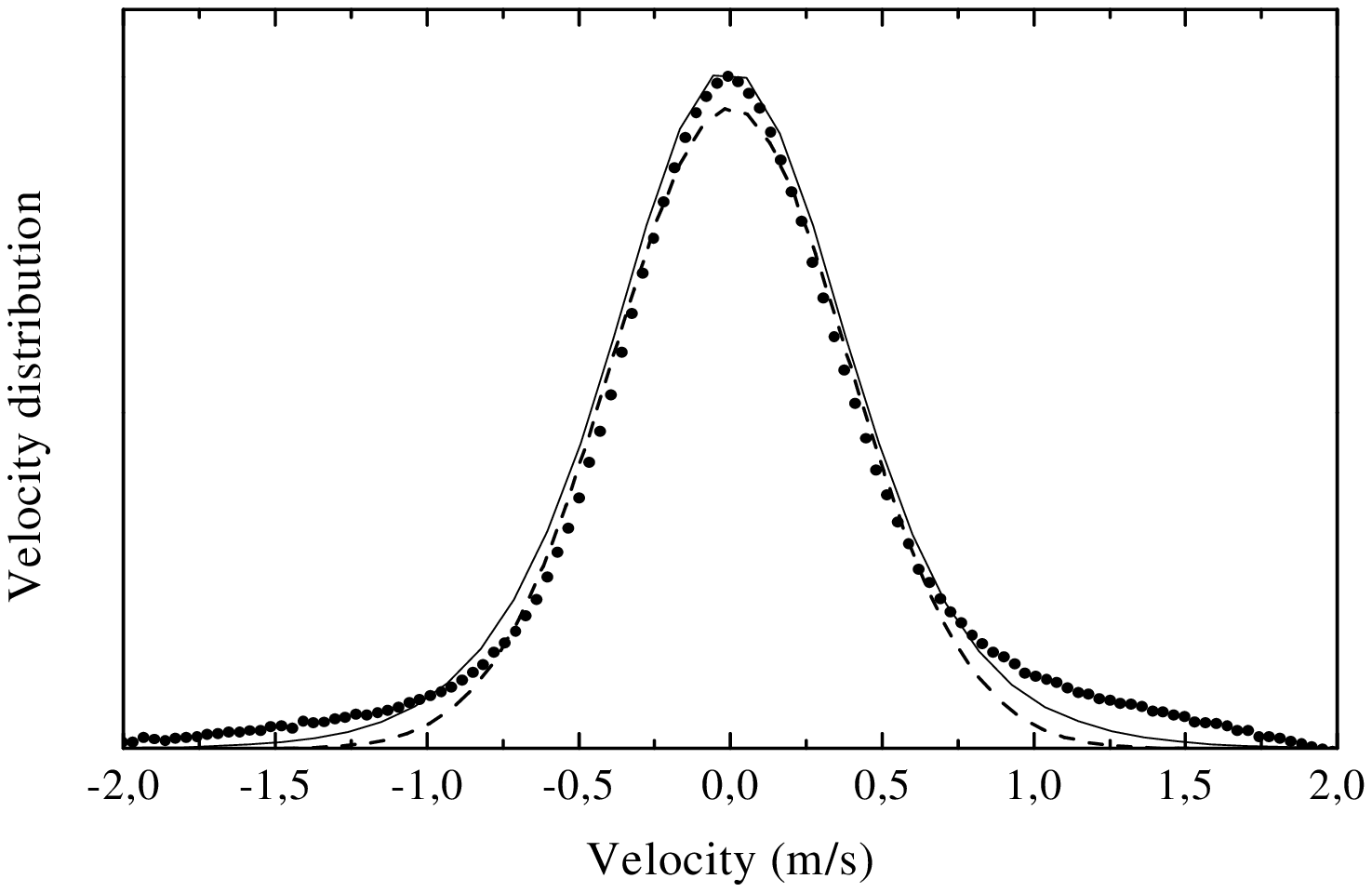}
\caption{Velocity distribution obtained at $s_0=0.08$ and
$\Delta=-1$. Circles : experiment. Solid line : Gaussian fit.
Dashed line : Monte-Carlo simulation with $r_s=7.5\%$ and
$\tau_s=20 \,\mu s$ (see text).} \label{TransProf}
\end{figure}

Figure \ref{TransProf} shows the longitudinal \emph{spatial}
distribution obtained for the longest ballistic time at
$\Delta=-1$ and $s_0=0.08$. It is two times broader than the
initial spatial distribution and essentially proportional to the
velocity distribution. We can then try to compare it to the
velocity distribution obtained with the MC simulation (solid
line). We have also plotted the Gaussian distribution expected
from standard Doppler theory (dashed line). The actual
non-Gaussian shape of the MC distribution is explained by the
transverse velocity dispersion ($\Delta v_{\perp}\approx 0.8
\,m/s$). For each \emph{fixed} transverse velocity, the
distribution is quasi-Gaussian with a width depending on
$v_{\perp}$ (see figure \ref{analytic} and discussions in Section
\ref{analytic_model}). For this simple reason, summing over the
transverse velocity distribution leads to a non-Gaussian
distribution. The non-Gaussian shape of the MC simulation matches
the experimental distribution better than the Gaussian one.

As a conclusion of this section, we again stress that we have
quantitatively explained both the behavior of the velocity
dispersion $\sigma_v$ as a function of laser detuning and
intensity and the observed non-Gaussian distributions. This puts
strong evidence on the fundamental role played by molasses
intensity defects in the cooling process.

\section{Cooling dynamics}
\label{dynamic_cooling}

\begin{figure}[]
\includegraphics[scale=0.5]{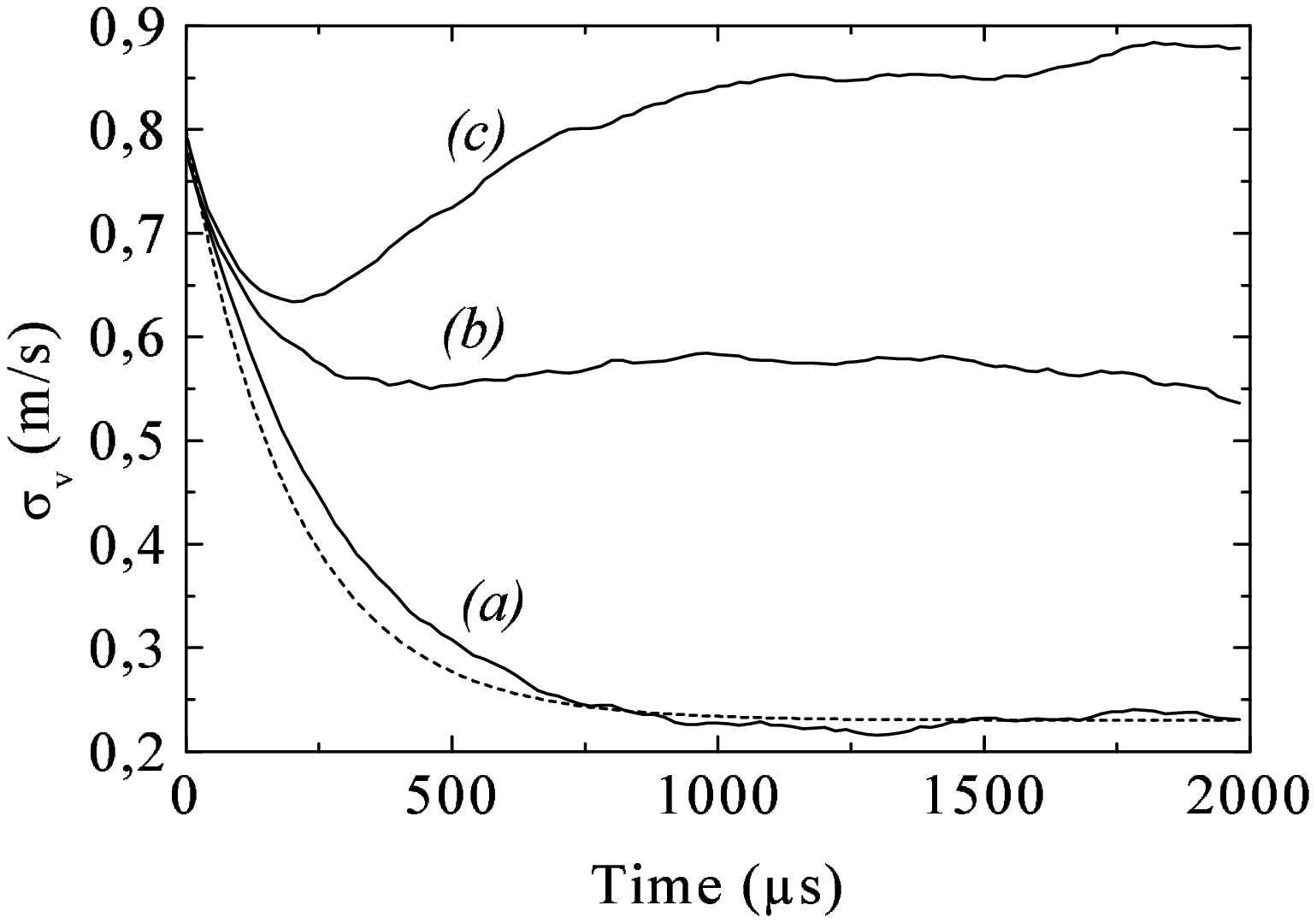}
\caption{Monte-Carlo simulations of the time evolution of the
velocity dispersion $\sigma_v$ for three characteristic
correlation times: (a) short correlation time ($\tau_s=1.25\,\mu
s$) ; (b) intermediate correlation time ($\tau_s=125\,\mu s$) ;
(c) long correlation time ($\tau_s=1250\,\mu s$). The transverse
velocity dispersion is $\Delta v_\perp=0.8m/s$, the mean
saturation per beam is $s_0=0.04$ and the saturation fluctuation
parameter is $r_s=9\%$. The dashed exponential decay is the bare
Doppler prediction (see text).} \label{Dyn_anal}
\end{figure}

\begin{figure}[]
\includegraphics[scale=0.5]{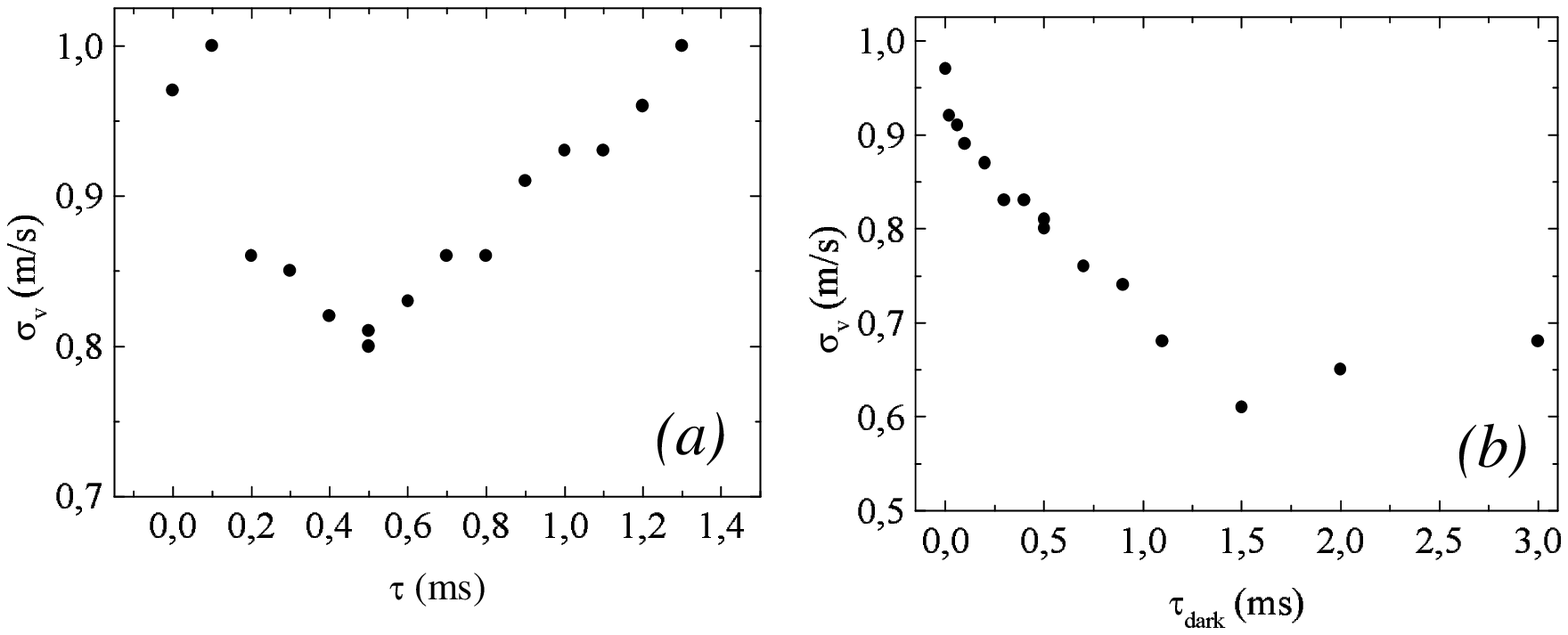}
\caption{(a) : Time evolution of the velocity dispersion
$\sigma_v$ after a dark time window of duration $\tau_{dark}=0.5
\,ms$ ; (b) MOT velocity dispersion $\sigma_v$ as a function of
the dark window duration. The dark window starts $0.5 \,ms$ after
the beginning of the cooling sequence (see text).} \label{Dyn_exp}
\end{figure}

Figure \ref{Dyn_anal} show the time evolution of $\sigma_v$ for
three different values of $\tau_c$ corresponding to the short,
intermediate and long correlation time limit. When $\tau_s$ is
short (or equivalently $v_{\perp}$ large), the corresponding curve
(a) displays an exponential-type behavior decaying to the Doppler
steady-state value. This curve is in agreement with the dashed
line Doppler prediction. When $\tau_s$ is long (or equivalently
$v_{\perp}$ small), the dynamics evidenced by curve (c) is more
complex. First $\sigma_v$ reaches a minimum value within a time
scale corresponding to the damping time $\tau_v$. Then, $\sigma_v$
increases again and reaches the steady state value
$\sigma_{\infty}$ predicted by (\ref{long c}). When the mean
velocity and the local intensity imbalance are uncorrelated in the
initial state, then this long correlation time behavior is generic
and does not depend anymore on the initial state. Because the
intensity correlation length $\xi_s$ is large, the heating
mechanism takes also some time to build up and the initial
velocity distribution starts first to shrink. This is easily
explained by considering the dynamics induced by (\ref{force}) for
times shorter than $\tau_s$ where transverse intensity
fluctuations are dynamically frozen. After a proper averaging over
intensities fluctuations, the following analytic expression for
$\sigma_v(t)$ is derived:
\begin{equation}
\sigma_v(t)^2=(\Delta
v^2+\sigma_{\infty}^2)\exp{(-2t/\tau_v)}+\sigma_{\infty}^2(1-\exp{(-t/\tau_v)})
\label{evol_comp}\end{equation} where $\Delta v$ corresponds to
the initial longitudinal velocity dispersion. the small time
expansion $t \ll \tau_v$ of this equation gives:
\begin{equation}
\sigma_v(t) \approx \Delta v \, (1-t/\tau_v)
\label{evol_court}\end{equation} clearly evidencing the velocity
spread narrowing at small times. We have experimentally tested
this specific behavior. Once the MOT is loaded, we have switched
off the cooling laser beams within a time window of variable
duration $\tau_{dark}$. This dark sequence must be long enough to
allow atoms to travel over transverse distances larger than the
correlation length $\xi_s$ of intensity fluctuations. Then, once
the cooling lasers are switched on again, any correlation between
the mean atomic velocity and the local intensity imbalance is
wiped out. We should thus observe the behavior predicted by curve
(c) in figure \ref{Dyn_anal}. The velocity dispersion $\sigma_v$
is measured after a time $\tau$ by the TOF technique (see section
\ref{Experimental Results}). In figure \ref{Dyn_exp}(a), we plot
$\sigma_v$ as a function of $\tau$ after a dark period
$\tau_{dark}=0.5 \,ms$. We reproduce nicely the corresponding
theoretical prediction. First we observe a decrease of $\sigma_v$
followed by an increase, up to the final value (which is here the
same as the initial value). We also checked that the dark period
has to be long enough, as evidenced in figure \ref{Dyn_exp}(b).
When $\tau_{dark}$ is short enough (less than $1 \,ms$),
decorrelation between the mean atomic velocity and local intensity
imbalances is not completed. When $\tau_{dark}$ is long enough
(larger than $1 \,ms$), complete decorrelation is achieved and the
dark period no longer plays any role in the cooling dynamics.

For practical reasons, these experiments were done with the MOT
({\it i.e.} on an 3D-cooling configuration in the presence of the
magnetic field gradient) whereas the MC simulations were done for
a 1D-cooling scheme. This major difference does not allow for a
quantitative comparison between theory and experiment. However,
the qualitative agreement is pretty good. This result suggests
that the dominant extra-heating mechanism in a MOT is the same as
in 1D-molasses.

Because the velocity distribution always starts by a compression
period, one could imagine a cooling strategy implementing suitable
repetitions of dark time windows to achieve Doppler-limited
cooling. This simple idea however is not easy to handle because
complete decorrelation between initial atomic velocities and
intensity imbalances during these dark windows is requested. As
the cloud cools down, the duration of the subsequent dark periods
has to be increased accordingly to maintain this decorrelation.
Unfortunately, as soon as the cooling time becomes very long,
spurious effects such as large cloud expansions, then have time to
set in.

\section{conclusions}

We have evidenced in this paper the important role played by
transverse spatial intensity fluctuations in 1D laser cooling of
zero spin groundstate atoms where no Sisyphus cooling can occur.
For intensity imbalanced molasses beams, the total radiation
pressure force decomposes, at small velocities, into a friction
force and a constant force. The latter is at the root of an
additional heating mechanism. These two forces essentially depend
linearly on the molasses intensities. In the presence of
sub-Doppler cooling mechanisms, the friction term becomes
intensity-independent whereas the constant force remains
proportional to the intensity \cite{lett, steane, wermer}. Hence,
with sub-Doppler cooling, the effect of intensity imbalances can
be arbitrarily small and, in turn, does not play any significant
role. This is completely different for Doppler cooling where this
effect remains dominant in most experimental cases.

The cooling steady-state reached by the atoms depends sensitively
on the ratio $\zeta$ between the correlation length of transverse
intensity fluctuations and the transverse distance travelled by
the atoms before reaching the steady-state. For small $\zeta$, the
transverse fluctuations can be modelled by a Langevin force
inducing an additional heating mechanism. For large $\zeta$, the
equilibrium state is reached at frozen molasses intensities. Atoms
at different transverse positions then probe all possible
intensities imbalance. This averaging procedure mainly affects the
final atomic velocity and induces an inhomogeneous broadening of
the Doppler velocity distribution. The impact on the final
temperature is more severe than in the small $\zeta$ regime. This
is evidenced by our experimental data which mostly lie in the
large $\zeta$ regime. It is however possible, in principle, to
reach the small $\zeta$ regime by appropriately reducing the
intensities of the cooling beams and get final temperatures closer
to the Doppler limit.

The cooling dynamics also exhibits an unusual behavior in the
large $\zeta$ regime : the atoms are first cooled down before
being heated up. This specific feature is also found in our
experiments, thus confirming the central role of transverse
intensity fluctuations in Doppler cooling.

\section{Acknowledgments}

This research is financially supported by the CNRS (Centre
National de la Recherche Scientifique) and the BNM ( Bureau
National de M\'etrologie) contract N$^{\circ}$ 03 3 005.

\section{Appendix : Derivation of the master equation}
\label{DerivMasterEq} The Fokker-Planck type equation
(\ref{FockkerP}) is obtained by starting with the master equation
(\ref{master}):
\begin{equation}
\mathrm{P}(v, t+\tau_s)=\int \dd s_1\,\dd s_2 \,\dd
v'\;\mathcal{P}(s_1)\,\mathcal{P}(s_2)
\;\mathrm{P}(v',t)\,\delta(v-v'-\frac{F(s_1,s_2,v')}{m}\tau_s)
\end{equation}
By Fourier transforming this equation with respect to $v$, we get:
\begin{equation}
\widetilde{\mathrm{P}}(q,t+\tau_s)=\frac{1}{\sqrt{2\pi}} \int \dd
s_1\,\dd s_2 \,\dd v'\;\mathcal{P}(s_1)\,\mathcal{P}(s_2)
\;\mathrm{P}(v',t) \; \exp[-iq(v'+\frac{F(s_1,s_2,v')}{m}\tau_s)]
\end{equation}
We plug now expression (\ref{force}) of the force in this equation
to find:
\begin{equation}
\widetilde{\mathrm{P}}(q,t+\tau_s)=\frac{1}{\sqrt{2\pi}} \int \dd
v' \; \widetilde{\mathcal{P}}[q\tau_s(a-\gamma_v v')]\;
\widetilde{\mathcal{P}}^*[q\tau_s(a+\gamma_v v')] \;
\mathrm{P}(v',t) \exp(-iqv') \label{ouf}
\end{equation} where the
star denotes complex conjugation and where
\begin{equation} \widetilde{\mathcal{P}}(u) = \int \dd s \,
\mathcal{P}(s) \exp(-ius) \end{equation} is proportional to the
Fourier transform of the molasses saturation distribution
$\mathcal{P}$. As discussed in section \ref{Orign_imba}, the
distribution $\mathcal{P}$ is a Gaussian with mean $s_0$ and
dispersion $\sigma_s$. A Taylor expansion of (\ref{ouf}) up to
second order in $\tau_s$ then leads to:
\begin{equation}
\tau_v \, \frac{\partial \widetilde{\mathrm{P}}(q,t)}{\partial t}
\simeq \frac{1}{\sqrt{2\pi}} \int \dd v' \,
\Big[i\,qv'-q^2\frac{{\tau_s}}{2\tau_v}[v'^2(1+2r_s^2)+2r_s^2\frac{a^2}{\gamma_v^2}\big]\;
\mathrm{P}(v',t)\,\exp(-iqv')
\end{equation} An inverse Fourier transform
then gives equation (\ref{FockkerP}) once we note that
$\sigma_{\infty} = \sqrt{2} r_s a/\gamma_v$ according to equation
(\ref{long c}).

\begin{figure}[]
\includegraphics[scale=0.5]{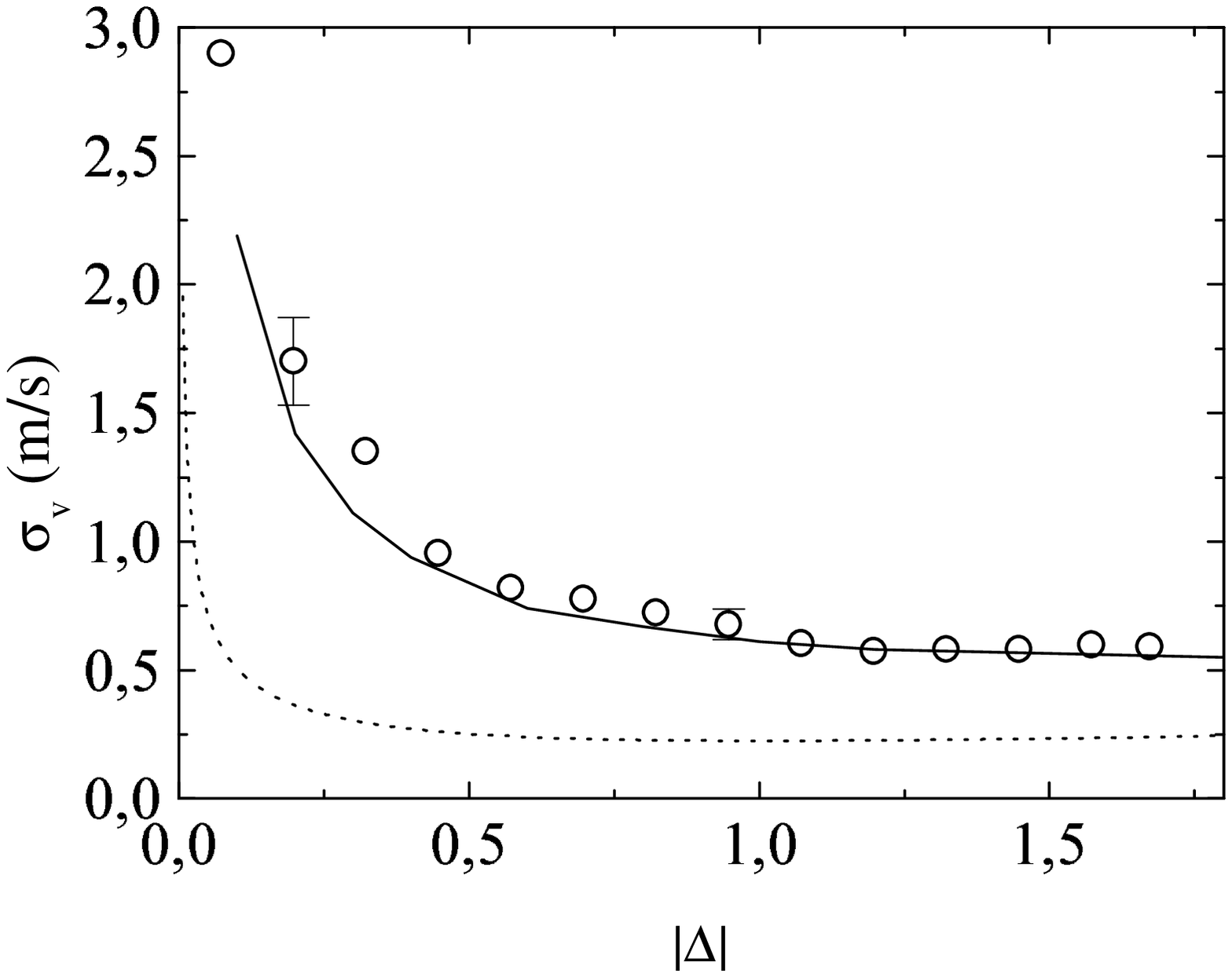}
\caption{Velocity dispersion $\sigma_v$ as a function of $|\Delta
|=2|\delta |/\Gamma$ for $s_0 = I/I_s=0.08$. The experimental data
(circles) are compared to the Doppler prediction (dotted line) and
to the Monte-Carlo simulation (solid line) at $r_s=9\%$ and
$\xi_s=60 \,\mu m$. As one can see the Doppler theory is
completely off while very good agreement is obtained with our
theoretical model (see text).} \label{delta}
\end{figure}

\end{document}